\documentclass[aps,prx,floatfix,twocolumn,superscriptaddress]{revtex4-2}
\usepackage{color}

\usepackage{graphicx,
            amsmath,
            braket,
            xcolor,
            xspace,
            lipsum}

\begin{document}

\title{Iterative assembly of $^{171}$Yb atom arrays with cavity-enhanced optical lattices}
\author{M.~A.~Norcia}
\thanks{M.~A.~N., H.~K.~and W.~B.~C.~contributed equally to this work.}
\author{H.~Kim}
\thanks{M.~A.~N., H.~K.~and W.~B.~C.~contributed equally to this work.}
\author{W.~B.~Cairncross}
\thanks{M.~A.~N., H.~K.~and W.~B.~C.~contributed equally to this work.}

\author{M.~Stone}
\author{A.~Ryou}
\author{M.~Jaffe}
\author{M.~O.~Brown}

\author{K.~Barnes}
\author{P.~Battaglino}
\author{T.~C.~Bohdanowicz}
\author{A.~Brown}
\author{K.~Cassella}
\author{C.-A.~Chen}
\author{R.~Coxe}
\author{D.~Crow}
\author{J.~Epstein}
\author{C.~Griger}
\author{E.~Halperin}
\author{F.~Hummel}
\author{A.~M.~W.~Jones}
\author{J.~M.~Kindem}
\author{J.~King}
\author{K.~Kotru}
\author{J.~Lauigan}
\author{M.~Li}
\author{M.~Lu}
\author{E.~Megidish}
\author{J.~Marjanovic}
\author{M.~McDonald}
\author{T.~Mittiga}
\author{J.~A.~Muniz}
\author{S.~Narayanaswami}
\author{C.~Nishiguchi}
\author{T.~Paule}
\author{K.~A.~Pawlak}
\author{L.~S.~Peng}
\author{K.~L.~Pudenz}
\author{D.~Rodr\'iguez~P\'erez}
\author{A.~Smull}
\author{D.~Stack}
\author{M.~Urbanek}
\author{R.~J.~M.~van de Veerdonk}
\author{Z.~Vendeiro}
\author{L.~Wadleigh}
\author{T.~Wilkason}
\author{T.-Y.~Wu}
\author{X.~Xie}
\author{E.~Zalys-Geller}
\author{X.~Zhang}
\author{B.~J.~Bloom\\ Atom Computing, Inc.}

\begin{abstract}
\noindent Assembling and maintaining large arrays of individually addressable atoms is a key requirement for continued scaling of neutral-atom-based quantum computers and simulators.  In this work, we demonstrate a new paradigm for assembly of atomic arrays, based on a synergistic combination of optical tweezers and cavity-enhanced optical lattices, and the incremental filling of a target array from a repetitively filled reservoir.  In this protocol, the tweezers provide microscopic rearrangement of atoms, while the cavity-enhanced lattices enable the creation of large numbers of optical  traps with sufficient depth for rapid low-loss imaging of atoms.  We apply this protocol to demonstrate near-deterministic filling (99\% per-site occupancy) of 1225-site arrays of optical traps.  Because the reservoir is repeatedly filled with fresh atoms, the array can be maintained in a filled state indefinitely.  We anticipate that this protocol will be compatible with mid-circuit reloading of atoms into a quantum processor, which will be a key capability for running large-scale error-corrected quantum computations whose durations exceed the lifetime of a single atom in the system.  

\end{abstract}
\maketitle

\section{Introduction}
Individually controlled neutral atoms provide a promising platform for quantum information processing and simulation, and expanding the size and control of these systems represent key challenges in the ongoing push to access regimes beyond the capabilities of classical simulation.  
Tweezer arrays and optical lattices have emerged as core technologies for optically trapping and manipulating cold atoms, each with complementary capabilities.  Optical lattices provide a well-defined potential landscape with features on the sub-wavelength scale, which can enable tight confinement for imaging and a low-disorder optical potential for simulations involving itinerant atoms \cite{gross2017quantum}.  Optical tweezers provide a means of selectively manipulating individual atoms \cite{kaufman2021quantum}.  Notably, tweezer arrays have gained prominence for deterministic assembly of nearly defect-free atomic ensembles \cite{kim2016in, barredo2016atom, endres2016atom} with up to hundreds of atoms \cite{scholl2021quantum, ebadi2021quantum}  and inter-atomic spacing suitable for Rydberg-mediated interactions between atoms \cite{wilk2010entanglement, Isenhower2010dem, madjarov2020high, scholl2021quantum, ebadi2021quantum, schine2022long, bluvstein2023logical}.  
Recently, optical tweezers and lattices have been combined to create programmable initial conditions in a Hubbard-regime system \cite{young2022tweezer, young2023atomic}, and to enable enhanced scaling of atom number for metrology \cite{young2020half} and quantum simulation\cite{tao2023high}.  

The merits of optical lattices can be further enhanced through the use of optical buildup cavities.  Optical cavities enhance the lattice depth by enabling the constructive interference of many overlapped reflections of laser light, in turn enabling the creation of large numbers of optical traps with sufficient depth to enable operations that require tight atomic confinement such as low-loss imaging.   \cite{bloom2014optical, park2022cavity}.  In our case, the use of cavity-enhanced lattices provides a roughly two-order-of-magnitude enhancement in power-efficiency for generating traps for imaging, when compared to optical tweezers (See appendix~\ref{app:comp}).
In this work, we combine the flexibility and control of optical tweezers with the scalable generation of deep traps afforded by cavity-enhanced optical lattices to demonstrate an iterative approach to creating large arrays of individually-controlled atoms.

\begin{figure*}[htb]
		\includegraphics[width=2\columnwidth]{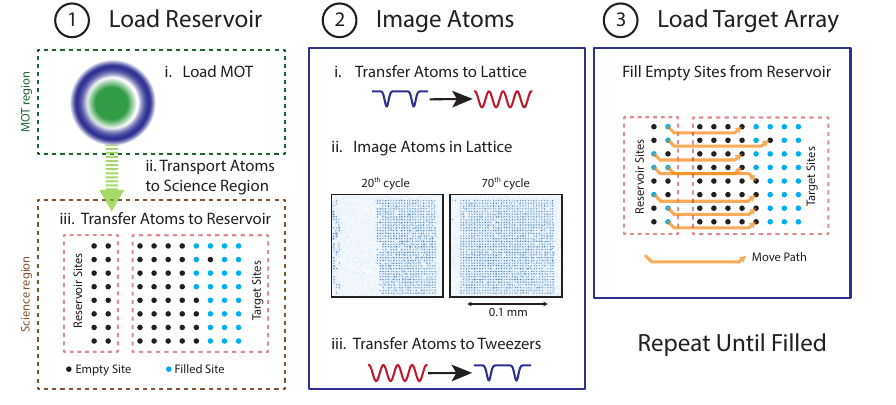}
		\caption{Conceptual illustration of repeated loading sequence.   (1) Load Reservoir:  The first step is to load an array of ``reservoir" tweezer traps with $^{171}$Yb atoms.  These atoms are collected in a spatially separated magneto-optical trap (MOT) (i), transported to the reservoir sites using a moving optical lattice (ii), and then transferred into the reservoir sites (iii).  During these steps, atoms loaded on previous iterations of the loading sequence are held in tweezers that form ``target" sites.  (2) Image Atoms: The second step is to determine the occupancy of reservoir and target sites.  This is done by transferring the atoms from the tweezers into a cavity-enhanced optical lattice (i), performing fluorescence imaging within the lattice, as shown after 20 and 70 iterations of loading (ii), and then transferring the atoms back into tweezers.  The cavity-enhanced lattice (described in section~\ref{sec:lattice}) allows for the efficient generation of large numbers of traps for imaging (see appendix~\ref{app:comp}).   (3) Load Target Array: The third step is to fill empty target sites with atoms from the reservoir.  The preceding image determines which sites are occupied, and a single mobile tweezer drags atoms from occupied reservoir sites to empty target sites.  This cycle can be repeated until the target array is filled.  }
		\label{fig:fig1}
\end{figure*}


Typically, optical tweezers can be used to assemble defect-free arrays of atoms by stochastically loading up to a single atom into each of a set of tweezers \cite{schlosser2001sub}, imaging the atoms to determine trap occupancy, and then rearranging atoms within the traps to create a deterministically occupied sub-array \cite{barredo2016atom, endres2016atom}.  Crucially, the number of atoms contained in the final ensemble with this approach is no greater than the number initially loaded.  Further, due to the initial stochastic loading, many extra sites are required to load a sufficient number of atoms to fill the target sub-array, especially when accounting for loss that occurs during the rearrangement process \cite{barry2023efficient} (though under certain conditions, near-deterministic loading can be achieved \cite{Brown2019Gray, jenkins2022yb}, and several initial arrays can be used to increase the number of available atoms \cite{Pause24}).  Recently, repeated loading of a ``buffer" array from an optical dipole trap ``reservoir" has demonstrated that one can decouple the filling of a six-site target array from a single loading of a cold reservoir \cite{pause2023reservoir}.  In this work, we extend this concept to repeated loading of a reservoir, from which we create a near-deterministically filled target array (typically 99\% occupancy) of over 1200 $^{171}$Yb atoms in 1225 sites. 

 Our approach allows us to maintain the filling of our atomic array for an arbitrary duration by replacing atoms that are lost.  In the near-term, this allows for relatively high data-rates for quantum simulation and computation with large system sizes, and could also be of benefit for optical clocks for high statistical precision with low dead-time \cite{dick1989local}.  

Ultimately, the ability to reload new atoms while maintaining both the presence and coherence of existing atoms (``mid-circuit reloading") will be a key capability for performing error-corrected quantum computations, where execution of an algorithm may take much longer than the lifetime of any given atom in the system. When combined with site-selective hiding \cite{lis2023mid, norcia2023midcircuit} and mid-circuit rearrangement techniques already demonstrated in a similar system \cite{norcia2023midcircuit}, we anticipate that this protocol will be compatible with mid-circuit reloading of atoms.  

As a means of maintaining a fully filled array, our approach represents an alternative to the interleaved use of two atomic species \cite{singh2022dual, singh2023mid}.  Unlike the two-species approach, the method presented here does not require simultaneous replacement of the entire array, and so may be expected to ease requirements for scaling to large arrays, and would not require inter-species gates \cite{anand2024dual} in order to replenish the array during a quantum computation.

\section{Repeated loading to build up a large array}\label{sec:repeated}

In this section, we provide a high-level description of our experimental apparatus and iterative loading protocol, with further details provided in the appendices.  A description of our cavity-enhanced optical lattices and imaging of atoms within the lattices can be found in section \ref{sec:lattice}.  The iterative protocol can be broken into three main phases, which are illustrated in figure~\ref{fig:fig1} and repeated until the array is filled: 1) Loading atoms into the reservoir tweezers, 2) Imaging the atoms by transferring atoms into the optical lattices and back, and 3) Filling empty sites in a target tweezer array with atoms from the reservoir.  

This sequence is enabled by an apparatus that is divided into two regions: a ``science" region and a ``MOT" (magneto-optical trap) region, which are separated by 30~cm  vertically.  Atoms are repeatedly collected in a MOT within the MOT region and transported to the science region using an optical lattice formed from counterpropagating 532~nm beams. In the science region, the atoms are accumulated in optical traps.  Importantly, each region has a distinct and static magnetic field environment: a magnetic-field gradient in the MOT region, and a large uniform field in the science region (500~Gauss, in the plane of the atom array -- the $x$ direction in figure~\ref{fig:lattices}a).  The static fields and spatial separation allow us to load atoms into the MOT without causing heating or loss of atoms in the science region.  This means that the MOT can be loaded while performing other operations in the science region \cite{norcia2023midcircuit}.  To further reduce the potential for scattered light from atoms in the MOT region being absorbed by atoms in the science region, we operate the MOT in a ``core-shell" configuration \cite{lee2015core} which has a dramatically lower scattering rate than a conventional sequential MOT (see appendix~\ref{app:mot}).  Details on the optical lattice used to transport atoms to the science region can be found in appendix~\ref{app:transport}.

Within the science region, several sets of optical traps facilitate the accumulation of atoms into a desired configuration.  These traps are either intensity maxima of optical tweezers, formed by tightly focusing laser beams through a high-numerical-aperture (NA = 0.65, field of view = 0.5~mm) microscope objective, or of co-located cavity-enhanced optical lattices, described in section \ref{sec:lattice}.  First, atoms are transferred from the transport lattice into a set of reservoir tweezers, where they are cooled and undergo light-assisted collisions (LACs) that ensure stochastic loading of zero or one atom per trap \cite{schlosser2001sub}.  In addition to the reservoir tweezers, the science region contains a square grid of 1225 ``target array" tweezers, where we accumulate atoms.  Importantly, we use a wavelength for the reservoir tweezers (483~nm) that does not shift the frequency of the narrow-linewidth $^1$S$_0$ $m_F = 1/2$ to $^3$P$_1$, $F = 3/2$ $m_F = 3/2$ transition used for cooling and LACs (referred to as a ``magic wavelength").  In contrast, the target array uses light at wavelengths (423~nm and 460~nm) that cause a large shift on the cooling and LAC transition, dramatically reducing the scattering rate and loss of atoms already loaded in the target array during cooling and LAC.  Further details on the reservoir and target array traps can be found in appendices~\ref{app:transport} and ~\ref{app:traps}.  

After loading the reservoir, the next step is to transfer the atoms into vacant target tweezers.  This requires identifying both which sites in the reservoir are occupied, and which sites in the target are vacant.  To do this, atoms from both tweezer arrays are transferred to the cavity-enhanced optical lattice, where they are imaged with low loss (described in section~\ref{sec:lattice}), and then transferred back into the tweezer arrays.  Based on the information gained from the image about which tweezer sites were unoccupied (either because they had yet to be filled or because a previously loaded atom had been lost), atoms are then moved from the reservoir into unoccupied target sites using a single ``rearrangement" optical tweezer (described in appendix~\ref{app:rearr}).  The cycle is then repeated until the target array is filled.  At this point, further operations may be performed on the atoms, and subsequent loading cycles are applied to maintain a filled array.

\begin{figure}[htb]
		\includegraphics[width=\columnwidth]{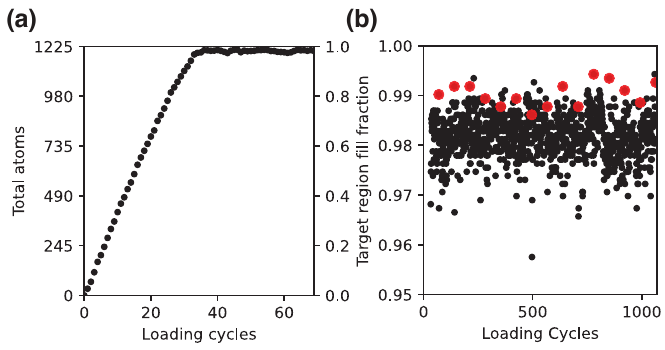}
		\caption{\textbf{(a)} Number of atoms and fill fraction as a function of loading iteration number during initial loading. \textbf{(b) The same quantities} while maintaining a filled array.  Black markers represent the atom number and fill fraction determined from images taken before the rearrangement step.  Red markers represent images taken after rearrangement on select iterations, which indicates the number of atoms available for computation (see main text for further details).  }
  
		\label{fig:continuous_load}
\end{figure}

Figure~\ref{fig:continuous_load}a shows an example of the atom number increasing as a function of load cycles, with 105 sites in the reservoir and 1225 sites in the target array.  At first, the atom number increases approximately linearly as roughly 45 atoms are transferred into the target array per cycle. The loading rate decreases slightly as the array fills, as some atoms from the reservoir are used to counteract loss.  Once the number of vacancies in the target array drops below the typical number of atoms in the reservoir, an equilibrium is reached with the vacancy fraction set by the per cycle probability of atom loss.  Figure~\ref{fig:continuous_load}b shows the fill fraction while maintaining a filled array by continuing to repeat the loading protocol.  

During loading, we only perform imaging prior to rearrangement.  These images are required to determine which sites need to be filled, and any additional images would lead to increased per-cycle loss.  The atom numbers and fill fractions inferred from these images are represented as black points in figure \ref{fig:continuous_load}.  Because these images are performed prior to rearrangement, the measured fill fraction is sensitive to any losses that have occurred since the equivalent image was performed in the previous loading cycle.    This fill fraction is typically 98\%, with the largest single loss contribution coming from vacuum loss during the typical 300~ms time between images due to our approximately 30~s vacuum lifetime.  Small variations in fill fraction may be attributed to drifting alignment of the tweezer arrays relative to lattices (see appendix~\ref{app:align}). We can add additional diagnostic images directly after rearrangement to determine the atom number available for computation (Shown as red markers in figure \ref{fig:continuous_load}b), and observe typical filling fractions of 99\%.  For these diagnostic images, the imperfect filling is dominated by vacuum loss during the shorter 150~ms interval typically allocated for rearrangement. Other sources of loss or potential loss are described in more detail in appendix~\ref{app:loss}, and include the handoff between the tweezer arrays and the lattice, and imaging loss and infidelity.

The number of atoms that we can currently load is limited by available laser power for the target array traps.  For larger tweezer arrays than those used here, the shallower tweezers lead to additional per-cycle loss when transferring atoms from the lattice to tweezers. This reduces the peak fill fraction, and if the array size is increased enough, leads to a situation where the number of atoms transferred from the reservoir on each cycle is insufficient to replace lost atoms.  The number of usable traps could be increased through the use of lasers with higher power or wavelengths with higher polarizability, as well as through more efficient optical paths and reduced optical aberrations.  

 As with all measurement-based rearrangement protocols, our filling fraction is limited by the atom loss that occurs between the image that identifies vacancies and the completion of the subsequent rearrangement step.  In our case, this  loss is dominated by background gas collisions that occur during rearrangement.  This could be improved either through lower vacuum pressure or by shortening the rearrangement sequence.  Several options exist for the latter:  A second iteration of rearrangement could be applied, condensing atoms into a slightly smaller sub-array \cite{Schymik2020enhanced}.  Because the number of vacancies to be filled would be lower and the distance that atoms must be transported to fill the vacancies would be shorter, this would enable a shorter rearrangement period, lower vacuum losses, and higher fill fraction.  Additionally, the rearrangement speed itself could be increased by improving optical aberrations on the rearrangement tweezer.


\section{Trapping and imaging atoms in a 3D cavity enhanced lattice}\label{sec:lattice}

\begin{figure*}[htb]
		\includegraphics[width=2\columnwidth]{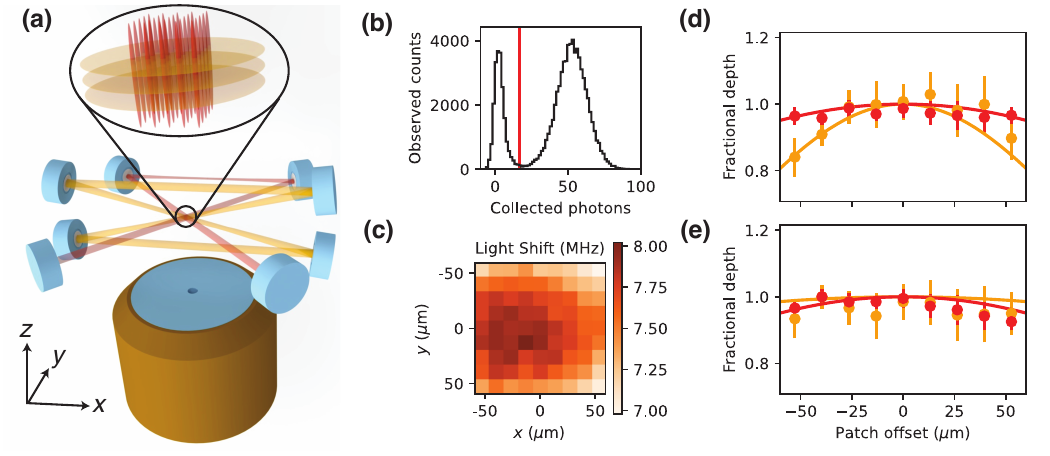}
		\caption{Cavity-enhanced optical lattices for rapid low-loss imaging.  \textbf{(a)} The intersection of two cavity modes provide three-dimensional confinement for atoms within the field of view of our high-numerical-aperture imaging system.  Confinement in the $x$ and $y$ directions is provided by a self-intersecting standing-wave cavity, while confinement in the $z$ direction is provided by a separate self-intersecting travelling-wave cavity mode.  Imaging light (not shown for clarity) is incident along the mode of the travelling-wave lattice.  \textbf{(b)} Observed counts during a 7~ms long image of single atoms within the optical lattice, showing well-resolved peaks for occupied and unoccupied sites.  For these conditions, per-image survival is 99.8(1)\%.  \textbf{(c)} Lattice homogeneity across the target array as characterized by light shifts on the $^1$S$_0$ $m_F = 1/2$ to $^3$P$_1$, $F = 3/2$ $m_F = -1/2$ transition.  Atoms within 13.4~$\mu$m squares are averaged together for statistics and display clarity.  \textbf{(d, e)} Light shift contributions from each cavity averaged over rows (d) and columns (e).  Red and orange markers and lines represent the measured and predicted profiles of the XY (red) and Z (orange) cavities, with the prediction centered on the array.  }
		\label{fig:lattices}
\end{figure*}

Rapid low-loss imaging of the locations of atoms is essential to our repeated loading protocol.  Because atoms are heated by the photon scattering required for imaging,  optical traps that are much deeper than the equilibrium atomic temperature established during imaging are required to prevent atom loss.  Optical tweezers represent a common method for achieving deep optical traps -- the tight foci can provide high intensities for moderate powers.  However, the total power requirement for a large array of tweezers presents scaling limitations due to limited available laser power.  Cavity-enhanced optical lattices offer an alternative path to achieving many deep traps.  While the laser intensity in a lattice is spread over a larger cross-sectional area than in optical tweezers both because it spans the regions between utilized sites, and because the Gaussian mode profile must be large compared to the size of the array to achieve uniform trapping, interference of many reflected paths in a cavity-enhanced lattice can enable more power-efficient generation of traps.   For our system, the power required to generate the 1225 imaging sites used in this work is roughly two orders of magnitude lower than the power that would be required to create the same number and depth of traps using optical tweezers of the same wavelength (see appendix~\ref{app:comp}).  This provides a promising path to scaling to much larger atomic ensembles.  

We employ two intersecting optical cavities to provide tight confinement in three dimensions, as shown in figure~\ref{fig:lattices}a.  
The first cavity -- which we call the XY cavity -- is a retro-reflected four-mirror cavity whose mode intersects itself at the location of the atoms.  The polarization is perpendicular to the plane of propagation, providing interference between the two crossed portions of the cavity mode \cite{sebby2006lattice}.  Compared to using two crossed non-interfering lattices \cite{heinz2021crossed, park2022cavity}, and assuming equal cavity finesse, this configuration provides an eight-fold enhancement in trap depth (defined as the potential barrier between adjacent lattice sites) per unit of total laser power.  The topology of this lattice configuration is also stable to translation of any mirror \cite{sebby2006lattice}, so the enhanced depth comes without penalty with respect to stability, and also benefits from fewer input-coupling paths and laser stabilization systems. The XY cavity provides tight confinement in the directions perpendicular to the high numerical aperture optical axis used for the tweezers and imaging.  

The second cavity, which we call the Z cavity, is a four-mirror travelling-wave cavity whose mode intersects itself at the location of the atoms \cite{cai2020monolithic} at 15 degree angles from the XY plane.  This provides confinement along the tweezer axis.  The polarization of the Z cavity is also oriented perpendicular to its intersection plane, enabling complete interference of the crossing modes (like the XY cavity, the contrast of this interference pattern is also robust to the longitudinal position of the mirrors).  

Each cavity has two highly reflecting mirrors, and two matched partially transmitting mirrors.  The finesse of the XY and Z cavities are 2900(100) and 3000(100) respectively,  as extracted from the measured cavity linewidths of 307(10)~kHz and 448(15)~kHz (photon lifetimes of 518(15)~ns and 355(10)~ns) and measured cavity free-spectral-ranges of 890.38~MHz and 1345.13~MHz.   Our observed finesse is consistent with the specified transmission of our mirror coatings and measurements in a test setup bound per-mirror losses to 50 parts-per-million, indicating a transmission-dominated cavity.  The finesses of the cavities were chosen to attain a large power-buildup while limiting conversion of laser frequency noise to amplitude noise that can occur in cavities with narrow linewidth. The design mode waists  of the XY, Z cavities at the location of the atoms are 268~$\mu$m and 183~$\mu$m respectively.  For imaging, we typically operate with XY (Z) trap frequencies near 160~kHz (50~kHz), corresponding to 330~$\mu$K (260~$\mu$K) deep traps. 

We image atoms in in the cavity using the narrow-linewidth (180~kHz) $^1$S$_0$ to $^3$P$_1$ transition, which simultaneously provides cooling and scattering of photons for detection \cite{saskin2019narrow, Huie2023rep, norcia2023midcircuit}.  We use a single  linearly polarized imaging beam, whose polarization is perpendicular to the magnetic field, and the propagation direction has projection onto both the X and Z directions.   The imaging beam is detuned several hundred kHz from the $F = 3/2$, $m_F = 3/2$ state, and has intensity approximately equal to the resonant saturation intensity of the imaging transition.  For applications like array assembly where we wish to determine the presence of an atom but not the state of its nuclear spin (the qubit we use for quantum information applications), we simultaneously apply optical pumping on the $^1$S$_0$ $m_F = -1/2$ to $^3$P$_1$ $F=1/2$, $m_F = 1/2$ transition with light incident along the magnetic field ($x$ direction in figure~\ref{fig:lattices}).  At our benchmark imaging parameters, we collect 50 photons in 7~ms, and distinguish occupied from unoccupied sites with a fidelity of $99.92^{+0.03}_{-0.04}$\%, determined from the consistency of classification in repeated images (see appendix~\ref{app:fidelity}).  Per-image atom loss in such repeated images is $2(1)\times10^{-3}$.  Spin-selective imaging requires us to omit the optical pumping, and rely on the combination of frequency and polarization selectivity within the $^3$P$_1$ manifold to create a large imbalance of scattering between the two nuclear spin states, as demonstrated in optical tweezers in refs.~\cite{Huie2023rep, norcia2023midcircuit}.  
The omission of the optical pumping does not modify the photon scattering rate appreciably, but does introduce a finite spin flip probability of of $4(1)\times10^{-3}$  for our typical imaging parameters, measured as an additional apparent loss in repeated images of $^1$S$_0$ $m_F = 1/2$.  These spin flips are likely due to the fact that the lattice polarization is perpendicular to the magnetic field, and so can induce spin-changing transitions within the $^3$P$_1$ manifold \cite{Huie2023rep}.  

In order to ensure uniform detuning of the imaging transition across the array and between different motional states, we operate with a trapping wavelength that has equal polarizability for the ground and excited states of the imaging transition.  For the $^1$S$_0$ $m_F = 1/2$ to $^3$P$_1$ $m_F = 3/2$ transition, we find such a condition near a wavelength of 783.8~nm  for lattice polarization perpendicular to the quantization axis.  For more details, see appendix~\ref{app:rearr}.

We characterize the homogeneity of the optical lattice potential using spectroscopy of the $^1$S$_0$ $m_F = 1/2$ to $^3$P$_1$ $F = 3/2$, $m_F = -1/2$ transition, for which the excited state has approximately 40\% higher polarizability than the ground state.  We infer the trap depth from the transition light shifts, measured using optical pumping from the $^1$S$_0$ $m_F = 1/2$ state, followed by spin-selective imaging (fig.~\ref{fig:lattices}c).  By varying the power in each cavity, we can extract their independent contributions, displayed averaged over rows and columns in figure ~\ref{fig:lattices}d,e.  Over our 115~$\mu$m square array, peak deviations are within 20\% for the Z lattice and 10\% for the XY lattice.  Because we operate with a magic wavelength condition for critical operations, these deviations  have only a small impact on imaging performance -- the minimum trap depth defines the limit on imaging speed for the whole array and excess trap depth leads to additional Raman scattering (discussed in appendix~\ref{app:loss}).

\section{Conclusions and outlooks}\label{sec:conclusion}
We have demonstrated an iterative method for assembling arrays of over 1200 individually controlled neutral atoms, suitable for quantum computation, simulation, and metrology.  Through this approach, we decouple the final size of the atomic ensemble from the number of atoms that can be loaded at once, and the need to create many tweezer traps with sufficient depth for loading and imaging atoms.  In particular, our reservoir and lattice traps are much deeper than the target tweezer traps, allowing for the use of large numbers of sites with moderate laser power.

For quantum computing applications, our iterative loading may be combined with recently demonstrated mid-circuit measurement techniques \cite{lis2023mid, norcia2023midcircuit} in order to achieve continuous mid-circuit refilling of array defects.  In such a protocol, a subset of atoms (``ancilla qubits") would be measured and potentially replaced while the rest of the atoms (``data qubits") maintain coherence as required to continue a quantum computation.  The roles of data and ancilla qubits may be alternated in order to allow replacement of any atom within the array.  In order to protect data qubits from decoherence from scattering imaging photons, tweezers operating at a wavelength that creates large light shifts to the imaging transition (as is the case for our target-array tweezers) can be applied to the locations of the data qubits, while all atoms are held in the lattice and illuminated with imaging light.   This shifts the data qubits off resonance with the imaging light, preventing decoherence, as we have demonstrated in a similar system with tweezer-confined atoms \cite{norcia2023midcircuit}.  Mid-circuit reloading also requires that light scattered from the MOT does not cause decoherence of the data qubits.  We have demonstrated that in our system, operating a broad-line MOT causes only a 0.03(2)/s decoherence rate for atoms in the science region \cite{norcia2023midcircuit}.   We expect that the core-shell configuration used here will have an even lower decoherence rate, due to a substantially reduced scattering rate (appendix~\ref{app:mot}).  By combining these capabilities with the continuous loading approach demonstrated here, it should be possible to enable the replacement of lost atoms within a circuit, which is a key capability for the execution of complex error-corrected circuits that may extend beyond the lifetime of an individual atom in the array.   For such a protocol to be successful, it is beneficial for all operations performed on atoms to have low rates of atom loss.  This is typically required for high-fidelity coherent operations, and with the techniques presented in this work and in reference~\cite{norcia2023midcircuit}, is possible for atomic readout as well.  

 Other approaches for mid-circuit measurement with neutral atoms include the use of two atomic species \cite{singh2023mid}, a readout zone \cite{bluvstein2023logical}, an optical cavity \cite{Deist2022mid}, or shelving of population to a state that does not couple to imaging light with either alkali \cite{graham2023mid} or alkaline-earth \cite{lis2023mid} atoms.  Extending these mid-circuit measurement protocols to mid-circuit reloading will face additional challenges.  If the reloading is to be performed quasi-indefinitely, there must be a way to replenish a depleted reservoir while maintaining coherence in the remaining atoms.  This has been demonstrated using the dual species approach \cite{singh2022dual}, though not with other methods.  As we describe above, our approach of separating the MOT loading region from the science region should also be compatible with mid-circuit reservoir refilling and we have previously demonstrated mid-circuit reloading from a reservoir in \cite{norcia2023midcircuit}.  

Unlike the dual species method, where the process of reloading ancilla atoms into the tweezers leads to the loss of atoms of that species already present, our approach requires only a small fraction of atoms to be reloaded on each reloading cycle.  This may ultimately provide an important advantage for  mid-circuit reloading in systems with large numbers of atoms.

\appendix
\section*{Appendices}

\renewcommand{\thesubsection}{\Alph{subsection}}
\begin{figure*}[htb]
		\includegraphics[width=2\columnwidth]{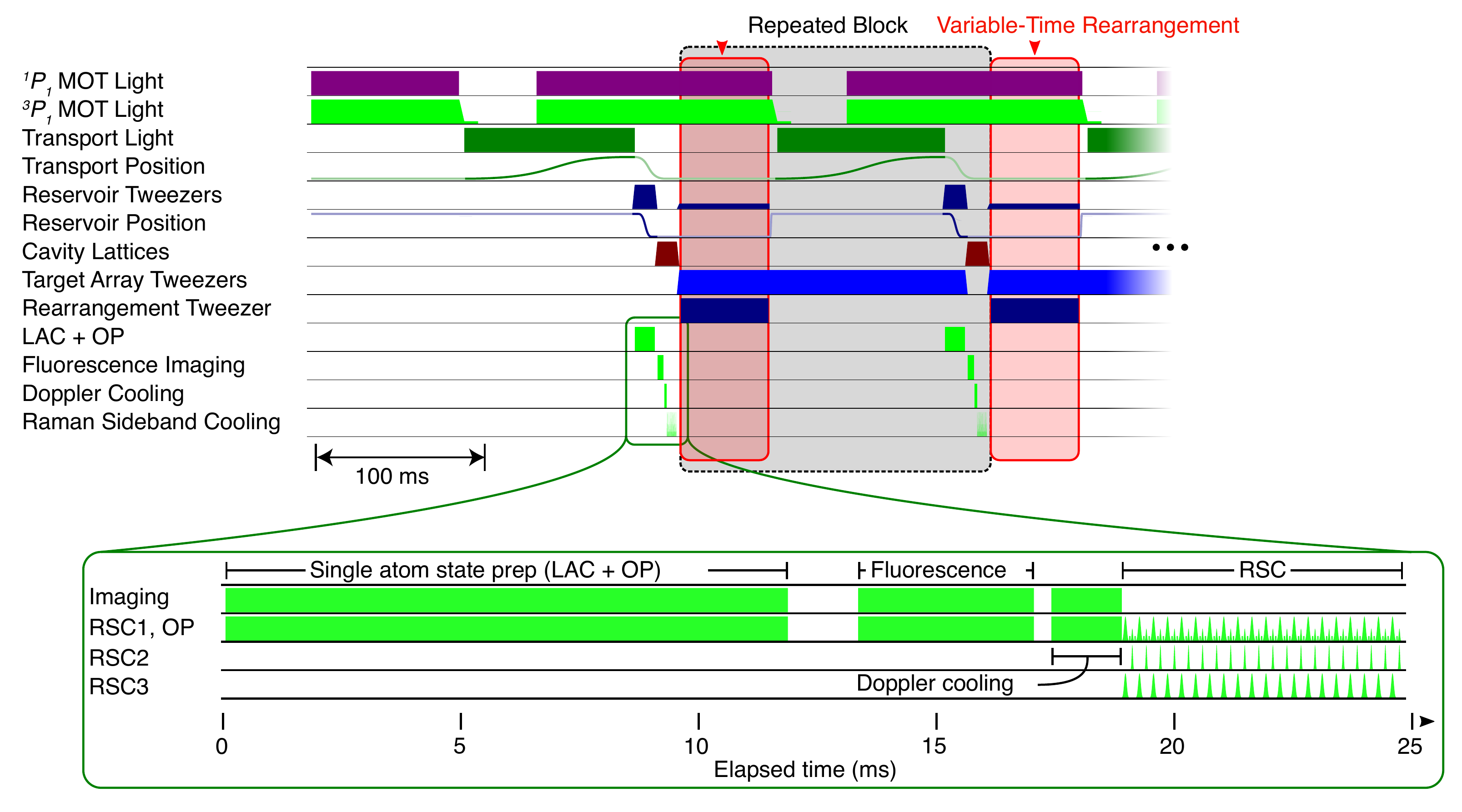}
		\caption{\textbf{(a)} Timing diagram for our continuous loading sequence.  See main text and appendices~\ref{app:mot}-\ref{app:rsc} for a detailed description of the referenced operations. The beam paths referenced in the call-out are illustrated in figure \ref{fig:rsb}.   The following acronyms are used in the figure: MOT: magneto-optical trap, LAC: light-assisted collisions, OP: optical pumping, RSC: Raman sideband cooling.  }
		\label{fig:sequence}
\end{figure*}
Appendices~\ref{app:mot}-\ref{app:align} describe Specific elements of our loading sequence.  An overview of the sequence with timing information can be found in figure~\ref{fig:sequence}.  Appendix~\ref{app:comp} provides a quantitative comparison of the power requirements for creating a given array of traps with optical tweezers versus cavity-enhanced lattices.  

\section{Core-shell MOT} \label{app:mot}
In order to minimize the potential for light scattered from the MOT to cause heating or decoherence of atoms in the science region, we operate with a core-shell configuration \cite{lee2015core}. In this MOT, each beam consists of a ring of 399~nm light near resonance with the broad linewidth $^1$S$_0$ to $^1$P$_1$ transition. This surrounds a central region with only 556~nm light that is near resonance with the narrow linewidth  $^1$S$_0$ to $^3$P$_1$ transition.    To create collimated hollow beams while minimizing the loss of laser power (compared to blocking the center), we use diffractive (HOLO/OR, DA-069-399-Y-A)-refractive (Thorlabs, AX252-A) axicon pairs. The diffractive axicon eliminates the round-tip effect seen with refractive axicons~\cite{Brzobohaty2008High}. The MOT is loaded from a precooled and collimated atomic beam provided by a commercial source (AOSense, Inc.). The shell is primarily responsible for capturing atoms from the incident beam, while the low Doppler temperature of the narrow transition used in the core enables cooling to several 10~$\mu$K.  Because atoms do not scatter light from the broad transition except during initial capture, the core-shell design poses reduced risk of generating scattered photons that may be absorbed by atoms in the target array -- the total scattering rate on the narrow transition is much lower than the broad transition, and the large magnetic field present in the science region dramatically suppresses reabsorption of photons via the narrow-line transition.  In our operating condition, the 399~nm photon scattering rate is about two orders of magnitude lower than the broad-line stage of a sequential MOT. We find that the core-shell MOT configuration increases the MOT loading rate by roughly a factor of two  (to $1.5 \times 10^6$ atoms per second), compared to a more standard sequential MOT configuration in a static magnetic field.   At this loading rate, we can collect enough atoms to saturate the filling of our reservoir within 80~ms.  This allows us to overlap the MOT loading in time with other operations performed in the science region.  

\section{Transport Lattice and loading of reservoir array} \label{app:transport}

From the core-shell MOT, atoms are loaded into a vertically oriented standing-wave lattice formed by counterpropagating 532~nm laser beams focused at the location of the MOT   with a beam diameter (1/e$^2$) of 75~$\mu$m. The typical depth of the lattice is 30~$\mu$K. Loading is achieved by overlapping the lattice with the center of the MOT, followed by reducing the detuning and intensity of the 556~nm light.  We then extinguish the MOT light and transport the atoms vertically by synchronously applying a frequency offset to the upward-going and downward-going lattice beams, and translating their foci over 100~ms with a profile calculated to minimize jerk \cite{norcia2023midcircuit}.  This brings the atoms 30~cm vertically into the vicinity of the target array, with several tens of thousands of atoms transported in a typical cycle. The typical survival during transport is 70~\%.   

We transfer the atoms from the transport lattice into a reservoir array formed by 483~nm tweezers (a magic wavelength for the $^1$S$_0$ to $^3$P$_1$ transition \cite{norcia2023midcircuit}).  To avoid disturbing atoms already loaded into the target array, this transfer takes place at a location displaced 170 $\mu$m horizontally from the target array.   No dissipation is required to transfer atoms from the transport lattice to the tweezers -- loading is achieved by simply increasing the depth of the tweezers  from zero to 250~$\mu$K over 1~ms while the lattice is overlapped and then decreasing the intensity of the  transport lattice to zero over 1~ms.  Once loaded, the reservoir array is translated  over 10~ms to be directly adjacent to the target array by changing the angle of a galvo mirror that is imaged onto the entrance pupil of the objective.   Compared to acousto-optic deflectors, the galvo mirror provides higher efficiency and a larger scan angle.  

\section{Target tweezer arrays} \label{app:traps}

The target tweezer array consists of a rectangular grid of tweezer spots formed with light at 459.5960(5)~nm wavelength, a magic wavelength for the $^1$S$_0$ to $^3$P$_0$ clock transition \cite{norcia2023midcircuit, hohn2023state}, and a set of tweezer spots formed by light at 423.31~nm, which is near resonance with a transition between $^3$P$_1$ to a higher-lying 6s8s~$^3$S$_1$ state.   All tweezers have waists of approximately 500~nm.  Both wavelengths provide large light-shifts to atoms in $^3$P$_1$, which prevents unwanted scattering from atoms loaded in these tweezer arrays, and can be used interchangeably for this work.  For this work, we operate with a 35x35-site target array with 3.3~$\mu$m spacing, comprised of two  side-by-side rectangular arrays each using one of the two previously mentioned wavelengths.   The typical trap depth in the target array is approximately 50~$\mu$K, which shifts $^3$P$_1$ by approximately 9~MHz in the 459~nm traps, and 50~MHz in the 423~nm traps (though this can be changed dramatically by varying the detuning from the $^3$P$_1$ to $^3$S$_1$ transition).  This requires approximately 200~$\mu$W delivered to the atoms per tweezer for the 459~nm traps and 100~$\mu$W per tweezer for the 423~nm traps. We have also operated with overlapped full-size arrays to achieve sufficient trap depth, and obtained similar results.  However, the overlapped condition is more sensitive to alignments, and so we use spatially separated arrays for all results presented here.

While translating the reservoir array adjacent to the target array (described in appendix~\ref{app:transport}), both arrays are illuminated with light resonant with the $^1$S$_0$ to $^3$P$_1$ $F=3/2$, $m_F=3/2$ transition.  This induces light-assisted collisions (LACs) between atoms in the reservoir tweezers, resulting in either zero or one atom in each reservoir tweezer with approximately equal probability \cite{schlosser2001sub}.  Importantly, the tweezer light illuminating the atoms in the target array prevents them from scattering the light used to induce collisions, which might otherwise cause loss of existing atoms in the target array. 

\section{Rearrangement from reservoir to target tweezer arrays} \label{app:rearr}

The rearrangement process relies on first identifying which sites of the reservoir and target arrays are occupied.  This is accomplished by transferring the atoms from the tweezer arrays to the cavity-enhanced lattice, and then imaging the atom locations.  Transfer is performed by turning on the lattice and simultaneously turning off the tweezers with linear intensity ramps lasting 1~ms.  The imaging leaves the atoms at a temperature of 20~$\mu$K, which is too warm to efficiently transfer back into the relatively shallow tweezers of the target array.  Following each image, we apply Doppler and Raman sideband cooling (RSC)  in the optical lattice to reach an average of $\sim$~0.1 motional quantum in each direction (appendix~\ref{app:rsc}).   The low temperatures obtained by RSC allow the atoms to be transferred efficiently into tweezer arrays of lower depth, which in turn allows for the generation of roughly 10 percent larger numbers of tweezers at fixed laser power compared to Doppler cooling alone.   Once cooled, the atoms are transferred back into the tweezer arrays, and a single optical tweezer derived from the same laser as the reservoir light is used to transfer atoms from filled reservoir sites into empty target sites.  This rearrangement tweezer is approximately 150~$\mu$K deep, which is chosen to balance the efficiency and speed with which an atom can be moved with disturbance of atoms already loaded into the array (see appendix~\ref{app:loss} and figure~\ref{fig:losses}). Our experimental control allocates a variable duration for rearrangement that depends on the number and length of moves to be performed, as well as the time required to compute the moves. The typical duration for the 1225-site arrays studied in this work is approximately 150~ms.  A possible future upgrade could involve using multiple rearrangement tweezers to move atoms in parallel, and thus speed up initial filling of the target array.

 Efficient imaging benefits from state-insensitive trapping for the $^1$S$_0$ to $^3$P$_1$ $F=3/2$, $m_F=3/2$ transition, which we find near 783.8~nm  for trap polarization perpendicular to the quantization axis.  This polarization-dependent magic condition (the polarizability of the excited state is 56\% greater for the orthogonal polarization of light) has a fractional deviation from state-insensitive conditions of 0.007/nm. Because the buildup cavities feature non-normal mirror reflections, they are highly birefringent and so provide polarization filtering, mitigating possible effects from drifting lattice polarization on the imaging transition frequency.

\section{Raman Sideband Cooling} \label{app:rsc}
\begin{figure*}[htb]
		\includegraphics[width=2\columnwidth]{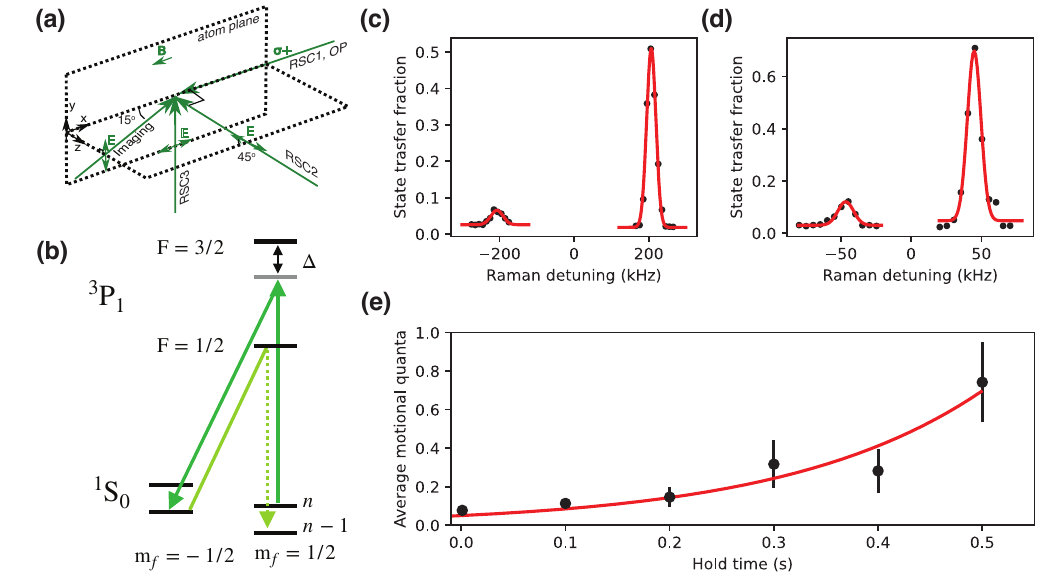}
		\caption{Raman sideband cooling (RSC).  \textbf{(a)} Diagram of relevant directions and polarizations.  RSC beams RSC1 and RSC2 are combined to cool the $z$ direction, while RSC1 and RSC3 are combined to cool the $x$ and $y$ directions.  Optical pumping (beam OP) is applied along the $x$ direction, which is along the applied magnetic field.  \textbf{(b)} Level diagram for Raman sideband cooling.  Raman transitions detuned by 35~MHz from $^3$P$_1$ $F=3/2$ $m_F = 1/2$ are driven from $^1$S$_0$ $m_F = 1/2$ to $^1$S$_0$ $m_F = -1/2$, while reducing the number of motional quanta along the addressed direction(s) from $n$ to $n-1$.  Optical pumping through  $^3$P$_1$ $F=1/2$ $m_F = 1/2$ returns atoms to $^1$S$_0$ $m_F = 1/2$ to enable further cycles. \textbf{(c)} Sideband spectra in the $x$, $y$ plane.  The sideband imbalance indicates an average number of motional quanta along the differential momentum vector of the RSC1 and RSC3 beams of $\bar{n}_{xy} = 0.08(1)$.  \textbf{(d)} Sideband spectra in the $z$ direction.  The sideband imbalance indicates an average number of motional quanta along the $z$ direction of $\bar{n}_{z} = 0.12(1)$.  \textbf{(e)} Heating in the lattice, as measured from the sideband imbalance in the $x$, $y$ plane, with a fit to an exponential growth profile, returning a time-constant of 190(40)~ms. }
		\label{fig:rsb}
\end{figure*}

Transferring atoms from the lattice into shallow tweezer traps with low loss requires that the atoms be much colder than the depth of the tweezer traps.  To accomplish this, we apply Raman sideband cooling (RSC) between the two ground nuclear spin states \cite{jenkins2022yb, lis2023mid} before transferring atom from the lattice to the tweezers.  For the motional Raman transitions, we use two pairs of beams oriented along the $x$ and $y$ directions, and along the $x$ and $z$ directions to provide cooling along all three directions (fig.~\ref{fig:rsb}a).  The Raman transitions transfer atoms from $^1$S$_0$ $m_F = 1/2$ to $^1$S$_0$ $m_F = -1/2$, state, and are detuned from the motional carrier transition by the appropriate trap oscillation frequency in order to reduce the motional state by one quantum (fig.~\ref{fig:rsb}b).   Optical pumping is provided by a beam oriented along $x$ and the magnetic field that addresses the $^1$S$_0$ $m_F = -1/2$ to $^1$P$_1$ $F = 1/2$, $m_F = 1/2$ transition.  

We perform 20 iterations of cooling, with each iteration consisting of a $\pi$ pulse on the red motional sideband for each pair, followed by optical pumping.  The pulse durations are 200~$\mu$s for the $xy$ pair and 100~$\mu$s for the $xz$ pair (tuned to the $z$ direction sideband), and 50~$\mu$s for optical pumping.  The total cooling sequence lasts 8~ms.  
In typical sequences, we apply a 2~ms Doppler cooling pulse prior to the RSC sequence using the same beam as for the imaging pulse, but with lower intensity and farther red-detuning.  The Doppler cooling results in 10~$\mu$K temperature in the cavity as measured by a release and recapture protocol~\cite{Tuchendler2008Energy}. This roughly corresponds to 1 motional quantum occupancy in the cavity.  Subsequent RSC reduces the average motional quanta in each direction to $\sim$~0.1, as evident from the sideband imbalance following cooling (fig.~\ref{fig:rsb}c, d).    

We measure heating in the $x$, $y$ plane while holding atoms in the lattice by performing Raman sideband spectroscopy to extract the average motional quantum number $\bar{n}_{xy}$ as a function of time.  We fit this quantity to a function representing exponential growth, as expected for parametric heating due to intensity fluctuations of the lattice \cite{savard1997laser}, and extract an exponential time-constant of 190(40)~ms.  This is likely due to conversion of laser frequency noise to amplitude by the cavity resonance, and could be improved by optimizing the lock of lasers to the cavity.

\section{Sources of Per-Cycle Loss} \label{app:loss}

\begin{figure}[htb]
		\includegraphics[width=\columnwidth]{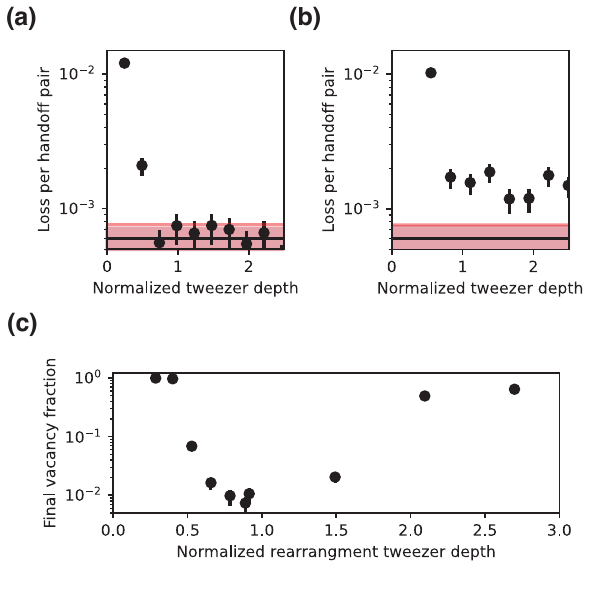}
		\caption{Characterization of per-cycle loss mechanisms.  \textbf{(a), (b).} Loss per pair of handoffs between the 459~nm (a) and 423~nm (b) tweezer arrays and lattice, measured by performing 25 handoffs with cooling in between.  The black line represents the estimated contribution from vacuum loss, with the red band representing day-to-day drifts in this value.  Tweezer depths are normalized to their default values, which correspond to roughly 50~$\mu$K.   The higher loss present in the 423~nm tweezer array is likely due to higher levels of optical aberrations compared to the 459~nm traps.   \textbf{(c)} Final vacancy fraction versus rearrangement tweezer power, normalized to the default tweezer depth of approximately 150~$\mu$K.  For low powers, atoms are not efficiently transferred from the reservoir to target array.  For too high of powers, we observe increased loss of loaded atoms due to the rearrangement tweezer passing nearby. }
		\label{fig:losses}
\end{figure}

Background gas collisions during the finite duration of loading cycles is the single largest source of loss in our system, contributing percent-level per-cycle loss.  Because our loading cycle is of variable length, and because the vacuum level in our system can fluctuate from day to day, it is difficult to estimate its exact contribution.  However, typical values for the loading cycle duration and vacuum lifetime are 300~ms and 30~s respectively, so 1\% represents a typical value for this loss.  In this section, we describe other loss sources that contribute to a lesser degree, especially if parameters are not carefully optimized.  In general, the loss mechanisms described below both limit the size of array we can ultimately load (as the reservoir must be large enough to replace lost atoms), and limit the final fill fraction of the array.  

As described in the main text, typical imaging loss is $2(1)\times10^{-3}$ where roughly 30~\% is the vacuum loss and the rest is the Raman scattering out of $^3$P$_1$ due to the trap light, and the discrimination infidelity is typically at or below the $10^{-3}$ level (see appendix~\ref{app:fidelity}).  The contribution from Raman scattering could be mitigated by applying repump lasers to the $^3$P$_0$ and $^3$P$_2$ states, though spin selectivity would be compromised in applications where that is required.   If an atom within the target array is lost during the image after scattering enough photons to be identified as present, the defect will not be filled during the subsequent rearrangement step, leading to a defect in the array.  If a site within the array is incorrectly identified as empty, an atom will likely be added to that site from the reservoir, and subsequently undergo lossy collisions with the original occupant.  Empty sites within the target or reservoir arrays that are mistakenly identified as full will lead to a previous defect not being repaired.  However, because defects are rare, this mechanism is less problematic.  We currently operate with a discrimination threshold that balances the correct identification of empty and full sites.  In a system where imaging losses and infidelity become dominant loss sources (a system with better vacuum), it may be advantageous to bias the threshold to minimize the more problematic forms of imaging error.  

Each image requires handing off atoms from the tweezer arrays into the lattice and back.  We find that this process is sensitive to the alignment of the two arrays on the scale of a single lattice site, to the depth of the tweezer arrays, and to the temperature of the atoms.  Atom temperature and alignment are described in appendix~\ref{app:rsc} and~\ref{app:align}, respectively.  We isolate the effect of trap depth on handoff for cold atoms and well-aligned arrays by performing 25 subsequent handoff pairs between tweezers and lattice in figure~\ref{fig:losses}a, b.  We perform our usual Raman sideband cooling (RSC) each time the atoms are in the lattice between handoffs.  For 459~nm tweezer depths below 75\% of our typical operating conditions, we observe significant loss.  Above this power level, we observe a constant loss  per handoff pair of 0.0006(1), which is consistent with data taken with the handoff omitted, and with our typical vacuum losses.  The 423~nm tweezers show a similar behavior, though with a slightly higher loss  per handoff pair of 0.0015(2) at high powers, likely due to worse alignment or optical aberrations.  If the tweezers are aligned to the trough of the XY cavity intensity, the loss can be as high as 0.02. 

In principle, moving new atoms in the target array may cause loss of existing atoms in the array, as the rearrangement tweezer moves near the occupied target sites.   In our rearrangement geometry, the moving tweezer passes 1.6~$\mu$m -- half the array spacing -- from the target sites.   We assess this possibility in figure~\ref{fig:losses}c by measuring the final fill fraction versus the depth of the rearrangement tweezer.  For a rearrangement tweezer that is too shallow, we observe a lower loading rate and so a lower final fill fraction.  For too deep a tweezer, we also see a reduction in the final fill fraction, as the rearrangement leads to loss of existing atoms.  Near our typical operating conditions, we observe a region with weak dependency of the fill fraction on the tweezer power.  

\section{Determining imaging fidelity}  \label{app:fidelity}
We use several methods to estimate our imaging fidelity.  The simplest (suitable for estimating the fidelity from limited data) is to generate a histogram of the number of collected photons (as shown in figure~\ref{fig:lattices}b) and fit a sum of two Gaussians to the peaks corresponding to background counts (indicating no atom in the imaged state) and signal counts (indicating an atom in the imaged state).  A threshold can then be defined to distinguish the presence or absence of an atom in the imaged state, and the fidelity estimated by the normalized area of the two Gaussians on the incorrect side of the threshold.  From this method, we estimate a typical imaging fidelity of 99.95\%.  While this method is convenient as it allows the fidelity to be estimated from datasets used for other purposes, it is not rigorous because the signal and background peaks may not be well-modeled by Gaussians.  

To provide a more rigorous calibration of the imaging fidelity, we perform many images in a row.  We identify misidentification events as instances when the qubit state identified in an image differs from the state in both the images that precede and follow it.  We perform this characterization in a stochastically loaded array (50\% filling), which evenly samples the possibility of false positive and negatives.  We divide the number of misidentification events by the total number of atom images (6570, excluding first and last images, as they cannot meet the criteria for misidentification) to extract a fidelity of $99.92^{+0.03}_{-0.04}$\%.  For our chosen threshold, the infidelity is evenly balanced between false positives and negatives, to within our statistical uncertainty.  

\section{Alignment of arrays}  \label{app:align}

\begin{figure}[htb]
		\includegraphics[width=\columnwidth]{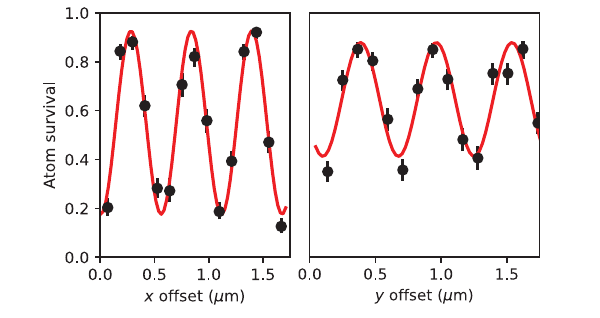}
		\caption{Alignment of the tweezer arrays to the cavity lattice is performed by monitoring atomic survival after repeated handoffs, while scanning the position offset of the tweezer array.  For the representative alignment scan of the 459~nm tweezer array shown here, we perform 25 handoff pairs between the tweezers and lattice, with no cooling in between.  Data is averaged over the full utilized field of view -- the presence of a visible fringe in the average image indicates that the sites across the array can be aligned simultaneously.  }
		\label{fig:array_align}
\end{figure}

Precise alignment of the different optical potentials is critical for the performance of our repeated loading protocol.  The alignment of the target tweezer arrays to the optical lattice is particularly sensitive, as misalignment here can lead to increased handoff loss that can limit both the largest array that can be loaded, and the final fill fraction of the array.  

Alignment of the arrays involves first matching the spacing and tilt of the lattice -- quantities which do not drift appreciably over time -- and then at a higher frequency aligning the $x$ and $y$ offsets.  We obtain the correct spacings and tilts by populating the lattice with atoms transferred from an expanded but sparse version of the reservoir array.  We image the atoms in the lattice to determine the lattice grid, and then in the reservoir tweezers to determine required corrections to the tweezer array.  We then use a camera whose imaging system is corrected for chromatic shifts between the different tweezer wavelengths to register the target array tweezers to the reservoir tweezers.  

We then optimize the translation alignment by performing repeated handoffs between the tweezers and the lattice and scanning the position offset of the tweezer arrays.  We observe a periodic modulation in the atomic survival (fig.~\ref{fig:array_align}), which we fit to determine the optimal alignment.  This process is repeated as necessary to ensure alignment of the arrays  (typically once per hour).

\section{Comparison of cavity-enhanced lattice to tweezers}  \label{app:comp}
Cavity-enhanced optical lattices and optical tweezers represent two alternative methods to generating large number of deep optical traps.  Here, we compare the efficiency of the two approaches in the context of generating large arrays of traps for atoms separated by distances appropriate for individual addressing and Rydberg interactions.  A third method -- using high-power lasers to generate deep lattices without the use of cavity enhancement -- can be inferred from the cavity analysis with a build-up factor of 1.  

For concreteness, we consider a square two-dimensional grid of $N$ lattice sites, separated by a distance $a$, trapped in either optical lattices or tweezers.  In the case of lattices, the grid is located at the intersection of several lattice beams, which can either be configured to interfere with each-other or not.  We define the depth $U$ as the amount of energy a classical atom would take to escape its local confinement -- the barrier between adjacent sites for the lattice, or the total depth for the tweezers -- and neglect gravity because its effect is negligible for traps deep enough to be suitable for rapid imaging.  For a cavity-enhanced lattice, the depth of the deepest traps is given by:
$$U_{l} = 2 \alpha P \Lambda I / \pi w^2$$

\noindent Here, $\alpha$ is the atomic polarizability.  $P$ is the laser power delivered to the cavity. $\Lambda = 2 F \epsilon/\pi$ is the power buildup factor. $F$ is the cavity finesse and $\epsilon  = T_{in}/(L_{tot} + T_{tot})$ quantifies the impedance matching to the cavity.  $T_{in}$ is the transmission of the input mirror, $T_{tot}$ is the total round-trip transmission of all mirrors in the cavity, and $L_{tot}$ is the total additional loss.  $w$ is the mode waist of the cavity at the location of the atoms.   $I$ quantifies the degree of interference at the location of the atoms.  For a simple standing-wave cavity, $I = 4$.  For a self-intersecting standing-wave cavity like our XY cavity, $I$ = 16.  Note that the self-intersecting cavity is a factor of eight more efficient because power must be split between two simple cavities to provide two directions of confinement.  A self-intersecting travelling-wave cavity like our Z cavity has $I = 4$.  The scaling of required power with atom number $N$ comes from the required increase of $w$ with larger atom number.  

For tweezers, the trap depth can be approximated as:
$$U_{t} = 2 \alpha P / \pi w_t^2 N$$
\noindent Here, $P$ is the total power used to form the tweezers, $w_t$ is the mode waist of the tweezers (approximated as Gaussian beams).  


For our cavity parameters and the atomic polarizability for 783.8~nm ($\alpha = 0.79$~MHz/mW/$\mu \rm{m}^2$), $U_l/P = 0.20$~MHz/mW for the XY cavity, and $U_l/P = 0.10$~MHz/mW for the Z cavity. 
We have set $\epsilon = 0.96$ here to represent the fact that the loss in both cavities is dominated by equal transmission through an input and output mirror (the other two mirrors in each cavity are coated for much higher reflectivity) and the measured bound on per-mirror loss.  For comparison, we may consider a tweezer array of 1225 traps with a waist $w_t = 700~nm$, which is typical for tweezer systems operating with wavelengths near 784~nm.  In this case, $U_t/P = 0.00083$~MHz/mW.  For this example, our XY cavity requires 255 times less power than the tweezer array to achieve the same depth, while the Z cavity requires 132 times less power.  Note that to provide three-dimensional confinement between adjacent sites in the lattice with a given depth, we must provide that depth of trapping with both the XY and Z cavities.  We have also neglected the efficiency of delivering light to the cavities relative to tweezers, but in practice the efficiency can be similar for the cavities and tweezers.  This dramatic improvement in power efficiency is especially important for architectures that require or benefit from the use of magic-wavelength traps, as this can constrain the use of high-power lasers.

\bibliography{bib}

\end{document}